\documentclass[english,prb, reprint, subscriptaddress]{revtex4-2}
\usepackage[T1]{fontenc}
\usepackage[utf8x]{inputenc}
\usepackage{geometry}
\usepackage{color}
\usepackage{babel}
\usepackage{verbatim}
\usepackage{units}
\usepackage{textcomp}
\usepackage{amsmath}
\usepackage{graphicx}
\usepackage[unicode=true,pdfusetitle,
 bookmarks=true,bookmarksnumbered=false,bookmarksopen=false,
 breaklinks=false,pdfborder={0 0 0},pdfborderstyle={},backref=false,colorlinks=true]
 {hyperref}
\begin{document}
\global\long\def\vect#1{\overrightarrow{\mathbf{#1}}}%

\global\long\def\abs#1{\left|#1\right|}%

\global\long\def\av#1{\left\langle #1\right\rangle }%

\global\long\def\ket#1{\left|#1\right\rangle }%

\global\long\def\bra#1{\left\langle #1\right|}%

\global\long\def\tensorproduct{\otimes}%

\global\long\def\braket#1#2{\left\langle #1\mid#2\right\rangle }%

\global\long\def\omv{\overrightarrow{\Omega}}%

\global\long\def\inf{\infty}%

\title{A New Microscopic Look into Pressure Fields and Local Buoyancy}
\author{S. M. João}
\email{simao_meneses_joao@hotmail.com}

\address{Centro de Física das Universidades do Minho e Porto and Faculty of
Sciences, University of Porto, Rua do Campo Alegre 687, 4169-007 Porto,
Portugal}
\author{J. P. Santos Pires}
\address{Centro de Física das Universidades do Minho e Porto and Faculty of
Sciences, University of Porto, Rua do Campo Alegre 687, 4169-007 Porto,
Portugal}
\author{J. M. Viana Parente Lopes}
\email{jlopes@fc.up.pt}

\address{Centro de Física das Universidades do Minho e Porto and Faculty of
Sciences, University of Porto, Rua do Campo Alegre 687, 4169-007 Porto,
Portugal}
\begin{abstract}
The concept of local pressure is pivotal to describe many important
physical phenomena, such as buoyancy or atmospheric phenomena, which
always require the consideration of space-varying pressure fields.
These fields have precise definitions within the phenomenology of
hydro-thermodynamics, but a simple and pedagogical microscopic description
based on Statistical Mechanical is still lacking in present literature.
In this paper, we propose a new microscopic definition of the local
pressure field inside a classical fluid, relying on a local barometer
potential that is built into the many-particle Hamiltonian. Such a
setup allows the pressure to be locally defined, at an arbitrary point
inside the fluid, simply by doing a standard ensemble average of the
radial force exerted by the barometer potential on the gas' particles.
This setup is further used to give a microscopic derivation of the
generalized Archimedes's buoyancy principle, in the presence of an
arbitrary external field. As instructive examples, buoyancy force
fields are calculated for ideal fluids in the presence of: i) a uniform
force field, ii) a spherically symmetric harmonic confinement field,
and iii) a centrifugal rotating frame.
\end{abstract}
\maketitle

\section{Introduction}

Thermodynamics is an essential tool for understanding the world around
us, giving a self-contained but simple description of physical processes
involving many particles, but using a minimal number of variables.
For example, temperature and pressure are pivotal concepts in the
thermodynamics of fluids but are also intrinsically macroscopic, emerging
only if the number of particles is very large and maintained in some
state of equilibrium.

A connection between the basic laws that govern the mechanics of microscopic
degrees of freedom and their thermodynamic manifestation is provided
by the edifice of Statistical Physics\,\citep{reif_fundamentals_1965,Pauli1973,kardar2007}.
In its simplest description, the temperature of a fluid is seen as
a measure of the mean energy per particle, while pressure describes
the average momentum transferred from the jiggling particles to the
walls of its container by means of random elastic collisions.\,Despite
providing a seemingly clear-cut explanation of the microscopic foundations
of thermodynamics, several investigations on physics education\,\citep{loverude2002,loverude2003_1,heron2003_2,bierman2003,kautz2005_1,kautz2005_2,loverude2010}
have proved these concepts difficult to grasp for most physics/engineering
students, especially when required to provide insight on practical
situations that fall outside the scope of typical textbook examples\,\citep{loverude2003_1,heron2003_2}.
To address this problem, it becomes necessary for the community of
Physics teachers to implement new ways of explaining these concepts
and, in particular, devise a wider variety of examples that more clearly
connect to real-life situations, whilst retaining a simple enough
description that makes their physical interpretation as transparent
as possible.

In standard textbook treatments, pressure is defined as the average
force (per unit area) caused by microscopic particles that collide
elastically with the walls of the fluid's container.\,In this sense,
it is a global property of any fluid in thermodynamic equilibrium,
when confined to a finite volume. In contrast, many physical phenomena
are known to depend on a local definition of pressure, described as
a position-dependent scalar field that exists within the system. Atmospheric
physics\,\citep{andrews_2010} and hydrodynamics\,\citep{Landau1987}
are two important and paradigmatic examples where non-uniform pressure
fields are ubiquitous concepts. Nonetheless, a simpler example is
provided by the well-known phenomenon of buoyancy (Archimedes' Principle),
which relies upon the existence of unbalanced pressures acting on
different surfaces of an immersed solid body. This position-dependent
pressure is caused by the Earth's gravitational field, but generally
it may be due to any other external field that the fluid particles
experience. In the literature, a myriad of nonstandard derivations\,\citep{irving1950,kirkwood1949,boer1949,kac1973,leroy1985,vermillion1991}
and generalizations\,\citep{van_den_akker1990,wick1977,mistura1987}
of Arquimedes' principle can be found, some of which going beyond
the case of a non-interacting classical gas. Nonetheless, we have
not found pedagogical presentations in the literature that illustrate
how scalar pressure fields can emerge from a microscopic theory. The
goal of this work is to provide such a description based on a kinetic
theory of classical non-interacting particles, common knowledge for
an undergraduate student.

In this paper, we develop a simple microscopic description of non-uniform
pressure fields in ideal fluids, based solely on a standard canonical
formulation of Statistical Physics, as well as rather simple kinetic
theory arguments. We define the local pressure by introducing a fictitious
localized spherical ``soft-wall'' potential that serves as a testing
probe. This probe plays the role of a small barometer, whose volume
couples to the local pressure, defining a pressure field. Assuming
a thermalized gas, we derive a closed-form expression for the pressure
field in the presence of an arbitrary external potential and, with
it, describe the buoyant force acting upon a small body immersed in
this fluid. Finally, this generalized Archimedes' principle is made
concrete by the application to three simple examples: i) a fluid under
a uniform gravitational field, ii) a gas cloud confined by a three-dimensional
harmonic potential, and iii) a rotating two-dimensional fluid under
a centrifugal force.

\begin{figure}[t]
\vspace{-0.5cm}

\includegraphics[scale=0.17]{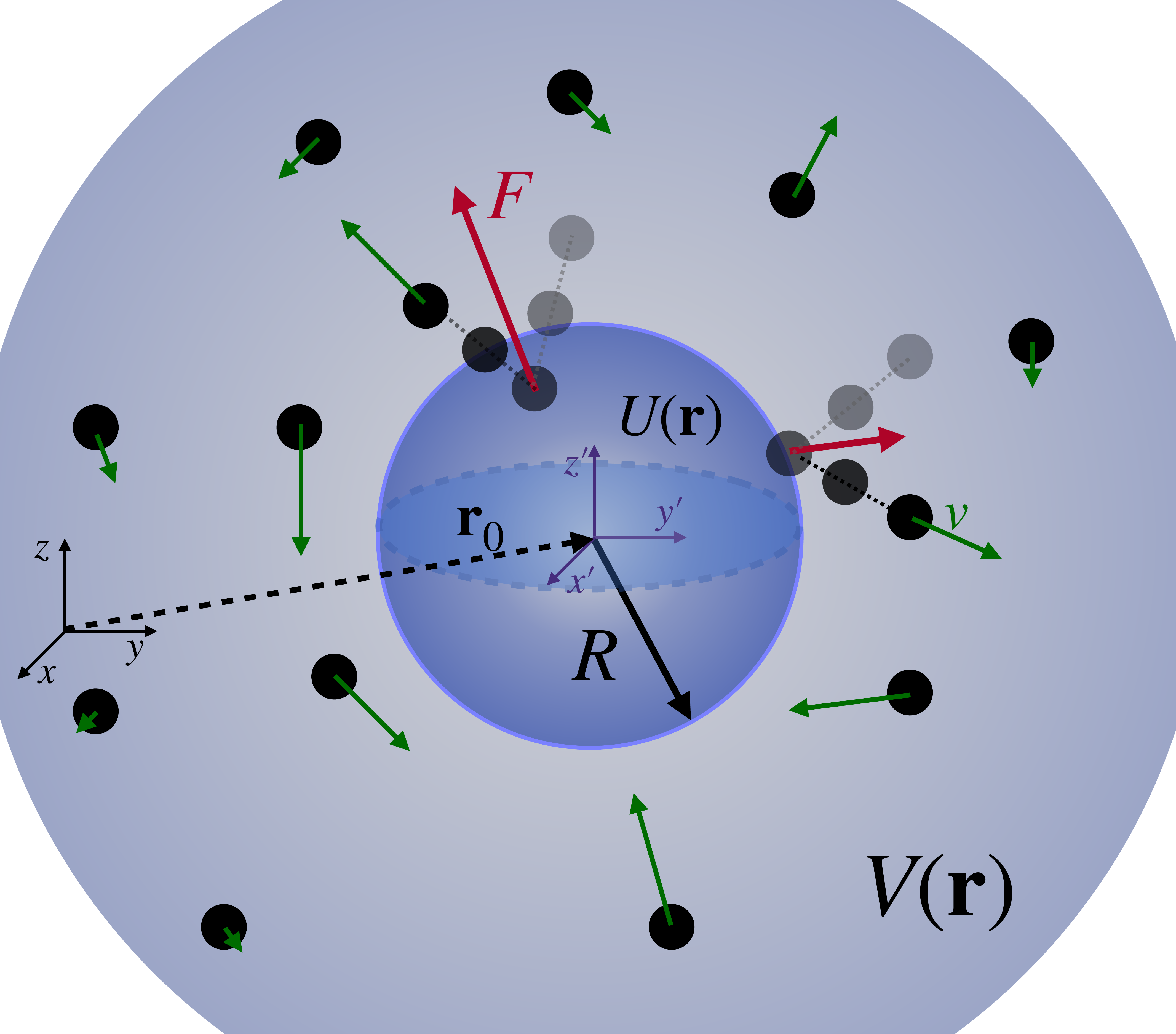}

\vspace{-0.2cm}\caption{\label{fig:sketch}Sketch of the physical situation.}

\vspace{-0.6cm}
\end{figure}

\vspace{-0.5cm}

\section{\label{sec:ModelofSphere}Microscopic Model of an Ideal Gas}

Aiming at a microscopic description of a fluid pressure field, we
start with a system having a large number $N$ of independent and
non-relativistic classical particles of mass $m$, that move across
a scalar potential $V\!\left(\mathbf{r}\right)$. This potential can
describe the confinement of the gas particles inside a container or
reflect an external force field acting globally upon the system (e.g.
a gravitational field). In order to have a consistent definition of
a position-dependent pressure field inside the system, one is forced
to introduce some kind of local probe. This is analogous to the more
common use of test electrical charges to define electromagnetic fields
in space. Therefore, we add to the Hamiltonian of the gas a short-ranged
perturbation potential $U_{\text{p}}\!\left(\mathbf{r}\right)$ representing
a small spherical probe of radius $R$ (as depicted in Fig.\,\ref{fig:sketch}).
The probe is assumed to be centered in an arbitrary but fixed position
$\mathbf{r}\!=\!\mathbf{r}_{0}$ (where the pressure is measured),
and $\mathbf{r}^{\prime}\!=\!\mathbf{r}\!-\!\mathbf{r}_{0}$ is taken
to be a position vector referenced to the center of the probe.

In the present context, we define the pressure at position $\mathbf{r}_{0}$
as the average momentum transferred from the gas particles to the
spherical probe, by means of collisions. To estimate this quantity,
it is useful to model the probe's surface, not as a rigid spherical
wall, but rather as a soft-wall potential which allows the particles
to slightly penetrate before being repelled outwards. This means that
any gas particle, at a distance $r^{\prime}\!=\!\abs{\mathbf{r}^{\prime}}$
from the probe's center will feel a spherically symmetric potential,

\vspace{-0.5cm}
\begin{equation}
U_{\text{p}}\!\left(\mathbf{r}^{\prime},R\right)\!=\!U_{0}f\!\left(\frac{r^{\prime}\!-\!R}{R}\!\right)\Theta_{\text{H}}\!\left(-\frac{r^{\prime}\!-\!R}{R}\right)\label{eq:ProbePotential}
\end{equation}
where $U_{0}$ measures the strength of the interaction, $\Theta_{\text{H}}(x)$
is the Heaviside step function and $f\!\left(x\right)$ is a continuous
and monotonous function that defines the overall shape of the soft-wall
potential, and behaves as $f\!\left(x\approx1^{-}\right)\!\approx\!-x$
close to the surface of the sphere (see Fig.\,\ref{fig.profiles}a).
Such a potential implies that the gas particles feel a roughly uniform
outwards radial force inside the sphere, but no force outside of it.
With the above definitions, the full Hamiltonian of the gas reads

\vspace{-0.4cm}

\begin{equation}
\mathcal{H}\!\left(\!\left\{ \mathbf{r}_{i},\mathbf{p}_{i}\right\} ,R\right)\!=\!\!\sum_{i=1}^{N}\!\left(\!\frac{\mathbf{p}_{i}^{2}}{2m}\!+\!V\left(\mathbf{r}_{i}\right)\!+\!U_{\text{p}}\!\left(\mathbf{r}_{i}\!-\!\mathbf{r}_{0},R\right)\!\right),\label{eq:Hamiltonian}
\end{equation}
where $(\mathbf{r}_{i},\mathbf{p}_{i})$ are the phase-space coordinates
of the corresponding particles. Note that, for simplicity, we have
considered the gas to consist of monoatomic particles that have no
(active) internal degrees of freedom.

\vspace{-0.5cm}

\section{Canonical Ensemble Formulation}

Having established our working microscopic model, we are now in position
to perform a standard Statistical Mechanical analysis of this non-interacting
gas. Concretely, we will work in the canonical ensemble (at a finite
temperature $T\!=\!1/\beta k_{\text{B}}$, where $k_{\text{B}}$ is
the Boltzmann constant) and calculate the canonical partition function
$Z$ starting from the basic Hamiltonian $\mathcal{H}$ defined in
Eq.\,(\ref{eq:Hamiltonian}). In terms of relative space coordinates,
$\mathbf{r}_{i}^{\prime}$, the partition function can be expressed
as the following $6N$-dimensional integral,

\vspace{-0.5cm}

\begin{equation}
Z\!\left(\beta,\!N\!;\!R\right)\!\!=\!\!\!\int\!\!d^{{\scriptscriptstyle \overset{\overset{\text{(3N)}}{}}{\,}}}\!\!\!\mathbf{r}^{\prime}\!\!\int\!\!d^{{\scriptscriptstyle \overset{\overset{\text{(3N)}}{}}{\,}}}\!\!\!\mathbf{p}\,\frac{e^{-\beta\mathcal{H}\left(\left\{ \mathbf{r}_{i}^{\prime}+\mathbf{r}_{0}\right\} ,\left\{ \mathbf{p}_{i}\right\} ,R\right)}}{N!h^{3N}},\label{eq:PartitionFunction}
\end{equation}
where $h^{-3N}$ is the conventional statistical measure in classical
phase space, while the $1/N!$ factor corrects the Gibbs paradox.
Since we are considering non-interacting particles, the partition
function naturally factorizes as a product of $N$ similar single-particle
terms. In addition, we can also easily integrate over the momenta
($3N$ independent gaussian degrees of freedom) which yields,

\vspace{-0.4cm}

\begin{equation}
Z\!\left(\beta,\!N\!;\!R\right)\!=\!\!\frac{1}{N!\lambda_{\text{\ensuremath{{\scriptscriptstyle \text{T}}}}}^{{\scriptscriptstyle 3N}}}\!\!\left[\int\!\!d^{{\scriptscriptstyle \overset{\overset{\text{(3)}}{}}{\,}}}\!\!\!\mathbf{r}^{\prime}e^{-\beta\left(V\left(\mathbf{r}_{0}+\mathbf{r}^{\prime}\right)+U\left(\!\mathbf{r}^{\prime},R\right)\right)}\!\right]^{N}\label{eq:PartitionFunction_Integrated}
\end{equation}
where $\lambda_{\text{T}}\!=\!\sqrt{\beta h^{2}/2\pi m}$ is the thermal
wavelength of a classical monoatomic particle. In the hard-probe limit
(i.e. $U_{0}\rightarrow\infty$), the integrand becomes exponentially
small inside the small probe's volume ($v$) and thus the integrals
of Eq.(\ref{eq:PartitionFunction_Integrated}) can be safely restricted
to its outside volume ($\overline{v}$), where $U\!\left(\mathbf{r}^{\prime}\!,R\right)\!=\!0$.
This further simplifies the expression of $Z$ as follows,

\vspace{-0.4cm}

\begin{equation}
Z\!\left(\beta,\!N\right)\!=\!\!\frac{1}{N!\lambda_{\text{\ensuremath{{\scriptscriptstyle \text{T}}}}}^{{\scriptscriptstyle 3N}}}\left[\int_{\overline{v}}\!\!\!\!\!\!\!\!d^{{\scriptscriptstyle (3)}}\!\!\mathbf{r}^{\prime}e^{-\beta V\left(\mathbf{r}_{0}+\mathbf{r}^{\prime}\right)}\right]^{N}\!\!.\label{eq:PartitionFunction_Integrated2}
\end{equation}
From the partition function of Eq.\,(\ref{eq:PartitionFunction_Integrated2}),
one can readily calculate the Helmholtz free energy --- $\mathcal{A}\!=\!-k_{\text{B}}T\ln Z$
--- and determine the thermodynamic properties of this system in
equilibrium. This quantity is expressed as

\vspace{-0.5cm}

\begin{align}
\mathcal{A}\!\left(\beta\!,\!N;\!v\right)\! & =\!-\frac{1}{\beta}\left[N\!\log\left(\int\!\!d^{{\scriptscriptstyle (3)}}\!\mathbf{r}^{\prime}e^{-\beta V\left(\mathbf{r}_{0}\!+\mathbf{r}^{\prime}\right)}\right.\right.\label{eq:HelmholtzFreeEnergy}\\
 & \qquad\!\left.\left.-\!\!\int_{v}\!\!d^{{\scriptscriptstyle (3)}}\!\mathbf{r}^{\prime}e^{-\beta V\left(\!\mathbf{r}_{0}\!+\mathbf{r}^{\prime}\right)}\right)\!-\!\log\left(N!\lambda^{{\scriptscriptstyle 3N}}\right)\!\right]\nonumber 
\end{align}
where we confined the whole contribution from the local pressure probe
to the second integral. Moreover, as the first integral in Eq.\,(\ref{eq:HelmholtzFreeEnergy})
extends over the entire volume of the system, it can be simplified
by changing the integration variable from the absolute position ($\mathbf{r}$),
to the position measured from the center of the probe ($\mathbf{r}^{\prime}$).
Finally, in the the limit of a point-like probe, the second integral
scales as $v$ and therefore one may expand the logarithm to first
order in $v$, yielding %

\begin{figure}[t]
\includegraphics[scale=0.24]{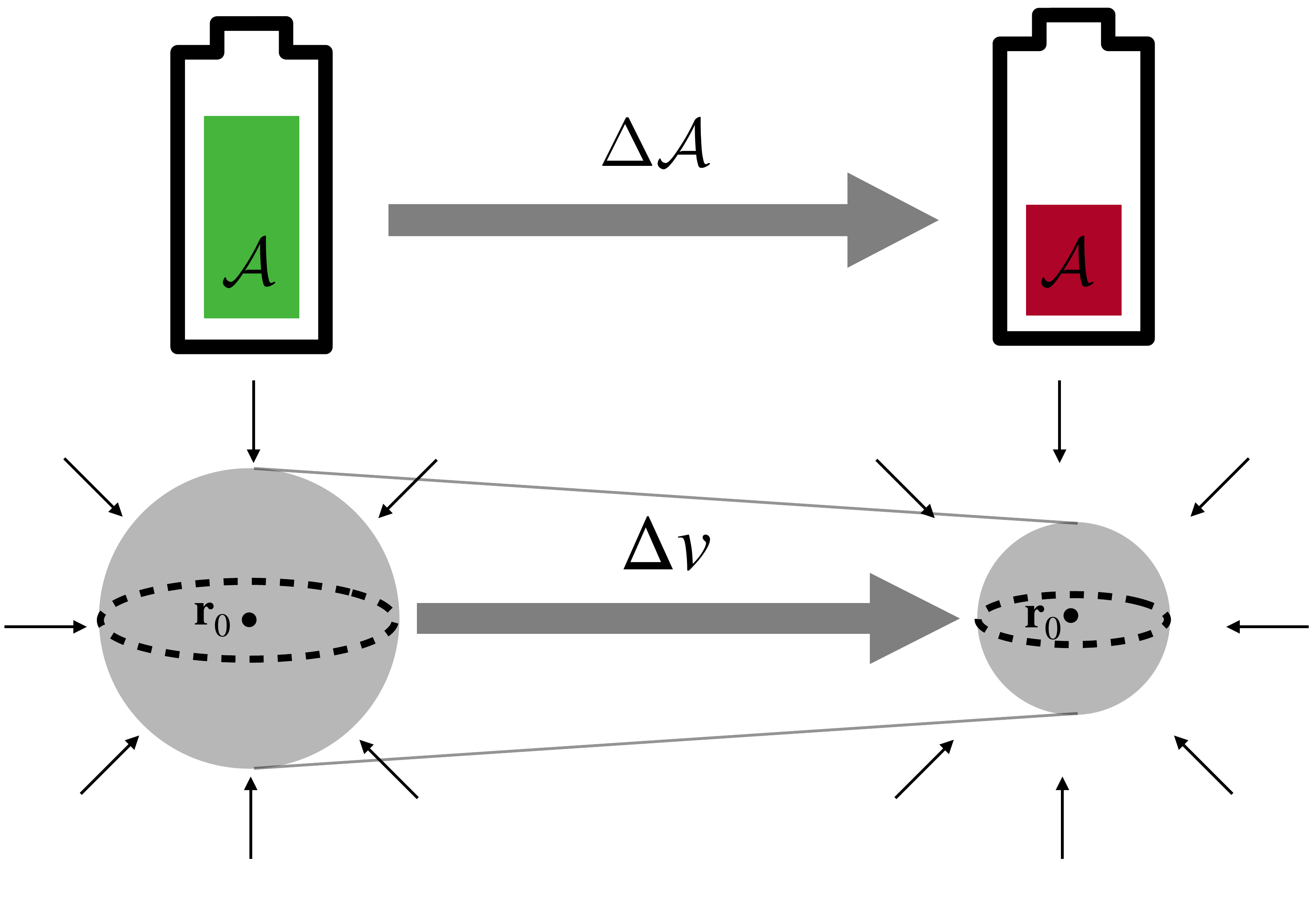}

\vspace{-0.3cm}

\caption{\label{fig:Cartoon-relating-changes}Cartoon relating changes in Helmholtz
free energy to variations in the value of the local probe. Note that
a decrease in the probe's volume results in a corresponding decrease
of the gas' Helmholtz free energy.}

\vspace{-0.3cm}
\end{figure}

\vspace{-0.3cm}

\begin{equation}
\mathcal{A}\!\left(\beta,N;v\right)\!=\!-\frac{N}{\beta}\log\left(\frac{V_{\text{eff}}}{N!\lambda^{3N}}\right)+v\frac{N}{\beta}\frac{e^{-\beta V\left(\mathbf{r}_{0}\right)}}{V_{\text{eff}}}.\label{eq:HelmholtzFreeEnergy-1-1}
\end{equation}
where we have defined the effective volume of the system as $V_{\text{eff}}\!=\!\!\int\!d^{3}\mathbf{r}e^{-\beta V\left(\mathbf{r}\right)}$.
In Eq.\,(\ref{eq:HelmholtzFreeEnergy-1-1}), the first term is simply
the Helmholtz free energy of a noninteracting monoatomic gas confined
inside a container of volume $V_{\text{eff}}$. In turn, the last
term is proportional to the volume of the small probing sphere, with
a position-dependent coefficient. Here, it is important to remark
that the thermodynamic definition of the pressure of a confined gas
is the volume derivative of $\mathcal{A}$, keeping $\beta$ and $N$
constant. Global pressure is the thermodynamic variable conjugate
to the whole volume. As will be microscopically derived, a similar
relation holds for the local pressure at point $\mathbf{r}_{0}$,
which will be the conjugate variable to the volume of a small probe
placed at that position. A cartoon of this is shown in Fig.\,\ref{fig:Cartoon-relating-changes}.

\vspace{-0.3cm}

\section{\label{sec:Non-uniform-pressure}Statistical Definition of a Fluid
Pressure Field}

The previous analysis was standard and, in principle, all equilibrium
thermodynamic quantities are encapsulated in the thermodynamic potential
$\mathcal{A}(\beta,N;v)$. Nevertheless, we intend to build a physical
reasoning that unambiguously defines the local pressure field, starting
from a familiar kinetic theory reasoning. Namely, we wish to derive
the average inwards force exerted on the walls of the spherical probe
by the colliding gas particles. By Newton's third law, this quantity
is simply the opposite to the force caused by the sphere's soft-wall
potential, $U\!(\mathbf{r})$, on any given particle penetrating its
surface.

Considering the $i^{\text{th}}$ particle, it will only feel a force
due to the probe when lying inside its radius. Then, the force will
point radially outwards and be given by $\mathbf{F}_{i}\!=\!-\nicefrac{\partial}{\partial r_{i}^{\prime}}U_{\text{p}}\!\left(r_{i}^{\prime},R\right)\hat{\mathbf{r}_{i}}^{\prime}$,
where $\hat{\mathbf{r}}_{i}^{\prime}$ is the unit vector corresponding
to $\mathbf{r}_{i}^{\prime}$. Following the previous reasoning, the
contribution to the inward pressure acting upon the spherical probe
is $\left(-\mathbf{F}_{i}\right)\!\cdot\!\left(-\hat{\mathbf{r}}_{i}^{\prime}\right)$
and, hence, the pressure due to the entire gas is simply

\vspace{-0.4cm}
\begin{equation}
P\!=\!\frac{1}{4\pi R^{2}}\left\langle \sum_{i=1}^{N}\mathbf{F}_{i}\!\cdot\!\hat{\mathbf{r}}_{i}^{\prime}\right\rangle ,\label{eq:PressureDefinition}
\end{equation}
where $\left\langle \cdots\right\rangle $ is a canonical ensemble
average. Meanwhile, the aforementioned equilibrium average can be
formally written as

\vspace{-0.4cm}

\begin{align}
\left\langle \!\sum_{i=1}^{N}\mathbf{F}_{i}\!\cdot\!\hat{\mathbf{r}}_{i}^{\prime}\!\right\rangle \! & =\!\!\!\!\int\!\!d^{{\scriptscriptstyle \overset{\overset{\text{(3N)}}{}}{\,}}}\!\!\!\mathbf{r}^{\prime}\!\!\int\!\!d^{{\scriptscriptstyle \overset{\overset{\text{(3N)}}{}}{\,}}}\!\!\!\mathbf{p}\sum_{i=1}^{N}\left[-\frac{\partial U_{\text{p}}\!\left(r_{i}^{\prime},\!R\right)}{\partial r_{i}^{\prime}}\!\right]\!\label{eq:EnsembleAveragedForce}\\
 & \qquad\qquad\qquad\times\frac{e^{-\beta\mathcal{H}\left(\!\left\{ \!\mathbf{r}_{i}^{\prime}\!+\mathbf{r}_{0}\!\right\} ,\left\{ \!\mathbf{p}_{i}\!\right\} ,R\right)}}{N!h^{3N}Z\!\left(\beta,N;R\right)}\nonumber 
\end{align}
where $Z$ is the partition function of Eqs.\,(\ref{eq:PartitionFunction})-(\ref{eq:PartitionFunction_Integrated2}).

Now, we can consider the rigid-probe limit ($U_{0}\rightarrow\infty$),
in which the function $(\partial U_{\text{p}}/\partial r')$ $\exp\left[-\beta U_{\text{p}}(r',R)\right]$
becomes highly sharply peaked in the vicinity of $r^{\prime}\!=\!R$.
In addition, as shown in Appendix.\,\ref{sec:Appendix:DiracDelta},
we also have $R(\partial U_{\text{p}}/\partial r')\!=\!-r^{\prime}(\partial U_{\text{p}}/\partial R)$
inside this integral, which leads to the relation%

\vspace{-0.4cm}
\begin{equation}
\frac{\partial U_{\text{p}}}{\partial r^{\prime}}e^{-\beta U_{\text{p}}}=-\frac{\partial U_{\text{p}}}{\partial R}e^{-\beta U_{\text{p}}}=\frac{1}{\beta}\frac{\partial}{\partial R}e^{-\beta U_{\text{p}}},\label{eq:Trick1}
\end{equation}
and using Eq.\,(\ref{eq:EnsembleAveragedForce}), it can be easily
shown that

\vspace{-0.4cm}

\begin{equation}
\left\langle \sum_{i=1}^{N}\mathbf{F}_{i}\!\cdot\!\hat{\mathbf{r}}_{i}^{\prime}\right\rangle \!=\!-\frac{\partial}{\partial R}\frac{1}{\beta}\log\!Z\left(\beta,N;R\right)
\end{equation}
and so the pressure exerted on the probe is simply

\vspace{-0.5cm}

\begin{equation}
P\!=\!\frac{\partial}{\partial v}\mathcal{A}\left(T,N;v\right),\label{eq:PressureThermodynamics}
\end{equation}
where $v\!\left(R\right)\!=\!\nicefrac{4}{3}\pi R^{3}$ is the volume
of the spherical probe. Note that Eq.\,(\ref{eq:PressureThermodynamics})
matches almost exactly the thermodynamic definition of pressure, as
being the change of the Helmholtz free energy upon an infinitesimal
change to the volume excluded by the spherical probe. The only difference
is in the sign. Typically, when a gas is contained within a macroscopic
volume $V$, any increase in this volume leads to a decrease in pressure.
Nevertheless, if the probe's volume $v$ increases in our setup, this
results in a diminishing available volume for the gas, thus increasing
the pressure. Despite the similarities to the usual case, it is important
to mention now the pressure is a function of the probe's position
$\mathbf{r}_{0}$ and, therefore, a locally defined quantity in the
limit when $v\!\to\!0$. Using the result obtained in Eq.\,(\ref{eq:HelmholtzFreeEnergy-1-1}),
our expression for the pressure in $\mathbf{r}_{0}$ is given by

\vspace{-0.5cm}

\begin{equation}
P(\mathbf{r}_{0})\!=\!\frac{N}{\beta V_{\text{eff}}}e^{-\beta V\!\left(\mathbf{r}_{0}\right)}\label{eq:Pressure_2}
\end{equation}
which can be easily expressed in terms of the equilibrium local density
of particles, $n\!\left(\mathbf{r}_{0}\right)$ (see Appendix\,\ref{sec:Appendix:-Density-of}),

\vspace{-0.5cm}

\begin{equation}
P\!\left(\mathbf{r}_{0}\right)\!=\!k_{\text{B}}T\,n_{\text{eq}}\!\left(\mathbf{r}_{0}\right).\label{eq:IdealGasLocal}
\end{equation}
This equation now takes a very recognizable form --- it is a local
version of the ideal gas law. Furthermore, we also remark that Eq.\,(\ref{eq:IdealGasLocal})
reduces to the (global) ideal gas law, in the absence of external
forces, but also reproduces the barometric formula, $P(x,y,z)\!=\!P_{0}e^{-z/z_{0}}$,
for a uniform gravitational field $V\!\left(z\right)\!=\!mgz$ and
Stevinus' law when expanded around $z\!=\!0$\,\citep{limaUsingSurfaceIntegrals2011}.

\vspace{-0.3cm}

\section{\label{sec:Microscopic-Derivation-of}Microscopic Derivation of the
Local Archimedes' Buoyancy Principle}

If the pressure inside a fluid is not uniform, then an immersed body
will be subject to a net force, historically known as buoyant force.
Physically, this happens when the pressures acting upon different
surfaces of the object are not balanced, and add up to a net force
that points in the direction of the lower pressure. In this section,
we will use our microscopic theory to derive this buoyant force in
the context of a general ideal fluid subject to an overall external
force field. Namely, we will focus on deriving an expression for the
buoyant force acting on the spherical probe, in the limit of hard-walls
and small volume, as a function of its location inside the fluid.

Evoking Newton's third law once again, the force exerted by the $i^{\text{th}}$
particle on the sphere is $\mathbf{-F}_{i}$, with the total force
being the ensemble average of the cumulative force due to all the
particles, i.e. $\mathbf{F}_{b}\!=\!\left\langle -\!\sum_{i=1}^{N}\!\mathbf{F}_{i}\right\rangle $.
Considering the gas to be in thermodynamic equilibrium, we arrive
at the following expression:

\vspace{-0.5cm}

\begin{align}
\mathbf{F}_{\!b}(\mathbf{r}_{0})\! & =\!\!\int\!\!d^{{\scriptscriptstyle \overset{\overset{\text{(3N)}}{}}{\,}}}\!\!\!\mathbf{r}^{\prime}\!\!\int\!\!d^{{\scriptscriptstyle \overset{\overset{\text{(3N)}}{}}{\,}}}\!\!\!\mathbf{p}\sum_{i=1}^{N}\left[\!\frac{\partial U_{\text{p}}\!\left(r_{i}^{\prime},R\right)}{\partial r_{i}^{\prime}}\hat{\mathbf{r}}_{i}^{\prime}\!\right]\!\label{eq:EnsembleAveragedForce-1}\\
 & \qquad\qquad\qquad\qquad\times\frac{e^{-\beta\mathcal{H}\left(\!\left\{ \!\mathbf{r}_{i}^{\prime}\!+\mathbf{r}_{0}\!\right\} ,\left\{ \!\mathbf{p}_{i}\!\right\} ,R\right)}}{N!h^{3N}Z\!\left(\beta,N;R\right)}.\nonumber 
\end{align}
Since the particles are non-interacting, Eq.\,(\ref{eq:EnsembleAveragedForce-1})
can be further simplified to

\vspace{-0.5cm}
\begin{equation}
\mathbf{F}_{\!b}(\mathbf{r}_{0})\!=\!-\frac{N}{\beta V_{\text{eff}}}\int d^{{\scriptscriptstyle (3)}}\mathbf{r}^{\prime}e^{-\beta V\left(\mathbf{r}_{0}\!+\!\mathbf{r}^{\prime}\right)}\hat{\mathbf{r}}^{\prime}\frac{\partial}{\partial r^{\prime}}e^{-\beta U_{\text{p}}\left(r^{\prime},R\right)}.\label{eq:AverageForce}
\end{equation}
Finally, we may consider the rigid-probe limit ($U_{0}\!\to\!\infty$),
such that the integral in the numerator of Eq.\,(\ref{eq:AverageForce})
becomes highly localized around the surface of the sphere. Then, we
can approximate $\nicefrac{\partial}{\partial r^{\prime}}\exp\left(-\beta U_{\text{p}}\left(r^{\prime},R\right)\right)\!\approx\!\delta\left(r^{\prime}\!-\!R\right)$
(see Appendix\,\ref{sec:Appendix:DiracDelta} for further details)
and perform a gradient expansion of $\exp\left(-\beta V\left(\mathbf{r}\right)\right)$,
i.e.

\vspace{-0.5cm}

\begin{equation}
e^{-\beta V\left(\mathbf{r}_{0}+\mathbf{r}^{\prime}\right)}\!\approx\!e^{-\beta V\left(\mathbf{r}_{0}\right)}\left[1\!-\!\beta\mathbf{r}^{\prime}\!\cdot\!\!\left.\mathbf{\boldsymbol{\nabla}}V\right|_{\mathbf{r}_{0}}\right].\label{eq:GradientExpansion}
\end{equation}
This way, the average force reduces to the much simpler form,

\vspace{-0.5cm}
\begin{align}
\mathbf{F}_{\!b}(\mathbf{r}_{0}) & \!=\!-\frac{N}{\beta V_{\text{eff}}}e^{-\beta V\left(\mathbf{r}_{0}\right)}\times\label{eq:Force}\\
 & \qquad\int\!\!\!d^{{\scriptscriptstyle (3)}}\!\mathbf{r}^{\prime}\delta\left(r^{\prime}\!\!-\!R\right)\!\left[\mathbf{r}'\!-\!\beta\mathbf{r}^{\prime}\!\cdot\!\!\left.\mathbf{\boldsymbol{\nabla}}V\right|_{\mathbf{r}_{0}}\right].\nonumber 
\end{align}
The physical interpretation of both terms in Eq.\,(\ref{eq:Force})
is very clear. The first term corresponds to a uniform pressure field
and integrates to zero. As expected, in the absence of external force
fields, the pressure is uniform across the system and does not yield
any buoyancy force. In contrast, the second term is proportional to
$\left.\mathbf{\boldsymbol{\nabla}}V\right|_{\mathbf{r}_{0}}$, which
measures the effect of the lowest order inhomogeneity in the field
at $\mathbf{r}_{0}$, i.e. a uniform pressure gradient. For evaluating
this term, the Dirac-$\delta$ can be directly employed by recognizing
that $\mathbf{r}^{\prime}\!=\!R\hat{\mathbf{r}}^{\prime}$, giving
a net force

\vspace{-0.5cm}

\begin{equation}
\!\!\!\!\mathbf{F}_{\!b}(\mathbf{r}_{0})\!=\!n_{\text{eq}}\!\left(\mathbf{r}_{0}\right)\!\!\int\!\!\!d^{{\scriptscriptstyle (3)}}\mathbf{r}^{\prime}\hat{\mathbf{r}}^{\prime}\delta\left(r^{\prime}\!\!-\!R\right)R\left[\hat{\mathbf{r}}^{\prime}\!\cdot\!\!\left.\mathbf{\boldsymbol{\nabla}}V\right|_{\mathbf{r}_{0}}\right]\!
\end{equation}
on the probe, where $n\!\left(\mathbf{r}_{0}\right)$ is the equilibrium
density of particles (as defined in Appendix\,\ref{sec:Appendix:-Density-of}).
Finally, by using the identity $\int d^{3}\mathbf{r}^{\prime}\hat{\mathbf{r}}^{\prime}\delta\left(r^{\prime}\!-\!R\right)\left[\mathbf{u}\!\cdot\!\hat{\mathbf{r}}^{\prime}\right]\!=\!\mathbf{u}\frac{4}{3}\pi R^{2}$,
valid for a general three-dimensional vector $\mathbf{u}$, we arrive
at the simple expression,

\vspace{-0.5cm}

\begin{equation}
\!\mathbf{F}_{\!b}(\mathbf{r}_{0})\!=\!n_{\text{eq}}\!\left(\mathbf{r}_{0}\right)v\left.\mathbf{\boldsymbol{\nabla}}V\right|_{\mathbf{r}_{0}},\label{eq:GeneralizedArchimedes-1}
\end{equation}
which shows the local buoyant force to be proportional to the local
value of the external force field. This expression allows us to define
a local buoyancy force field (per unit volume), $\mathbf{B}(\mathbf{r}_{0})\!=\!\mathbf{F}_{\!b}(\mathbf{r}_{0})/v$
and using Eq.\,(\ref{eq:IdealGasLocal}) we arrive at the generalized
Archimedes' Principle,

\vspace{-0.5cm}

\begin{equation}
\mathbf{B}(\mathbf{r})\!=\!-\boldsymbol{\mathbf{\nabla}}\!P(\mathbf{r}),\label{eq:GeneralizedArchimedes}
\end{equation}
which is valid for any non-uniform pressure field. This shows that
the pressure field is the \textit{effective scalar potential} associated
with the buoyancy force. Before moving on to some specific examples
of generalized buoyancy laws, we remark that $V\mathbf{B}(\mathbf{r}_{0})$
is the buoyancy force that acts on any immersed body of volume $V$,
independently of its shape. However, this is only true provided the
later is small when compared to the scale of spatial-variations in
the pressure field. For a larger immersed volume, its geometry starts
influencing the balance of pressures and Archimedes' principle strictly
breaks down. %

\vspace{-0.4cm}

\section{\label{sec:Examples}Some Illustrative Examples of Generalized Buoyancy
Laws}

\vspace{-0.2cm}

\noindent Previously, we have obtained a general expression for the
buoyancy force caused in a small body immersed inside an ideal fluid
under arbitrary external force fields. In this section, we will apply
the previous expressions to some specific examples, to further elucidate
the physical content of Eq.\,(\ref{eq:GeneralizedArchimedes}) and
provide non-standard examples of buoyancy. As announced in the introduction,
we will determine the pressure and buoyancy force fields for three
paradigmatic cases: \textit{i) }a fluid under a uniform gravitational
field, \textit{ii)} a gas cloud confined by a three-dimensional harmonic
potential, and\textit{ iii)} a rotating fluid inside a cylinder under
a centrifugal force. 
\begin{figure}[t]
\noindent \vspace{-0.4cm}

\includegraphics[scale=0.115]{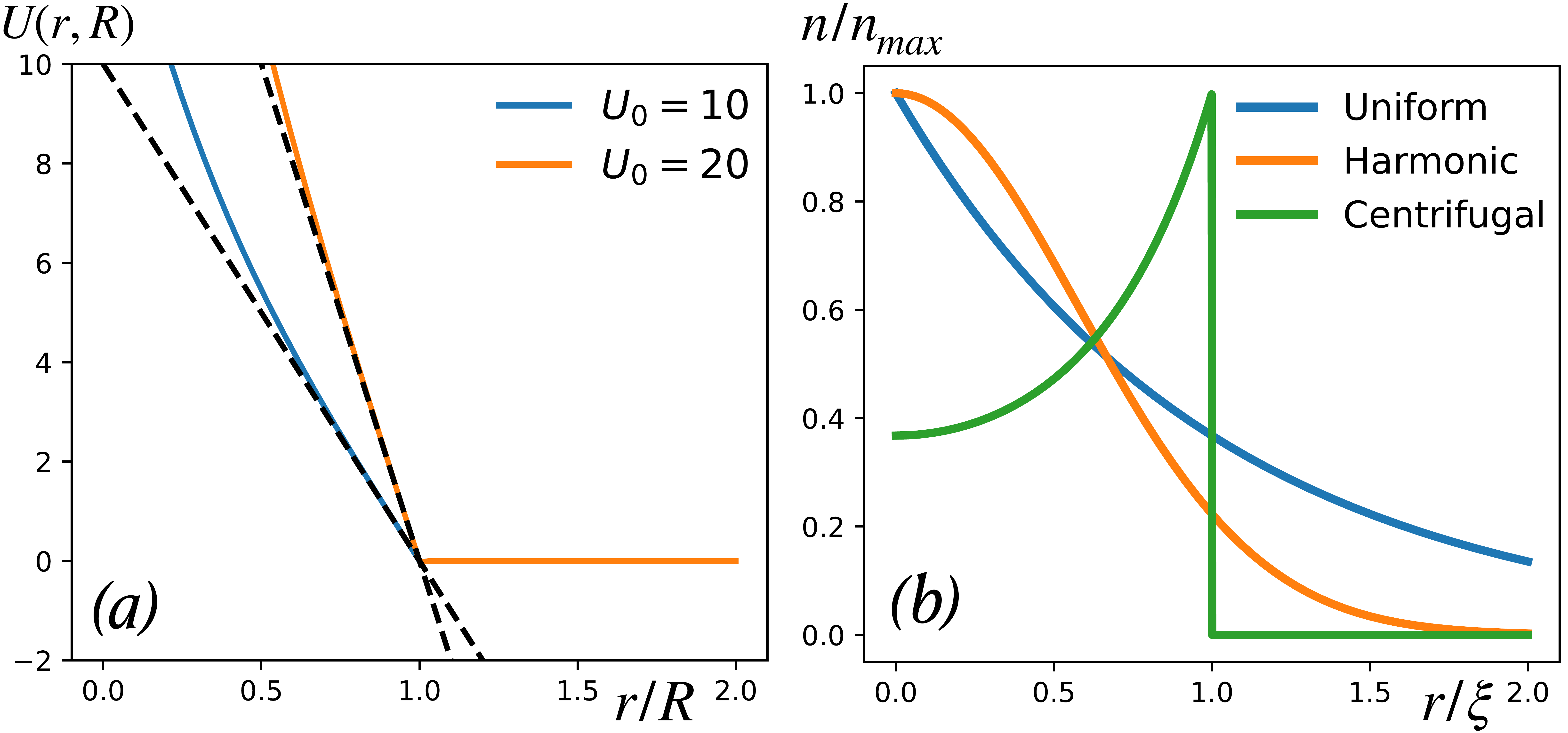}

\noindent \vspace{-0.4cm}

\caption{\label{fig.profiles}\textit{(a) }Soft-wall potential of the probe
for increasing values of $U_{0}$, approaching the rigid-probe limit.
The dashed lines represent the linear approximation to the potential
near the spherical surface, i.e $r\!=\!R$. \textit{(b)} Equilibrium
particle density profiles for the three examples considered. The curves
are normalized by their respective maxima and the distance is normalized
to the characteristic lengths (defined, for each case, in the main
text).}

\noindent \vspace{-0.4cm}
\end{figure}

\noindent \vspace{-0.8cm}

\subsection{Uniform Gravitational Field}

The simplest nontrivial case is that of a fluid on a uniform gravitational
field, i.e $V\!\left(\mathbf{r}\right)\!=\!V\!\left(z\right)\!=\!mgz$.
To simplify the analysis, we will confine the gas to lie inside a
column with a base area $A$ and an infinite height. By symmetry,
the equilibrium density of particles must only depend on the height
($z$) and is easily shown to decay exponentially with characteristic
distance $z_{0}\!=\!\left(\beta mg\right)^{-1}$ which is also the
average particle height. More precisely, the equilibrium particle
density is $n_{\text{eq}}\!\left(z\right)\!=\!(N/Az_{0})\exp(-z/z_{0})$
and the pressure follows the barometric law $P\!\left(z\right)\!=\!P_{0}e^{-z/z_{0}}$,
where $P_{0}\!=\!N/Az_{0}\beta$ is the pressure at ground level $z\!=\!0$.
Finally, the buoyancy force field can be easily obtained from Eq.\,(\ref{eq:GeneralizedArchimedes}),
yielding

\vspace{-0.4cm}
\begin{equation}
\mathbf{B}\left(z\right)\!=\!\frac{P_{0}}{z_{0}}e^{-z/z_{0}}\!\hat{\mathbf{z}}.
\end{equation}
This example can be made even more concrete by calculating the buoyancy
force which is exerted in a human being on the surface of the Earth.
For that purpose, we consider then average mass of an air molecule
($4.8\times10^{-26}\text{kg}$), a temperature of $300\text{K}$ and
a pressure at sea level of $10^{5}$ N/m\texttwosuperior{} (1 atm),
which estimates a typical variation length for the pressure field
to be $z_{0}\!\approx\!8772\,\text{m}$. Since $z_{0}$ is much larger
than the height of a person, the generalized Arquimedes' principle
applies and, given the average volume of a person ($6.2\!\times\!10^{-2}\!\text{m}^{3})$,
one concludes that a person would feel a total buoyancy force of $0.7$
N. This is obviously a minor effect that accounts for a reduction
of about 0.1\% in the perceived weight.

\vspace{-0.4cm}

\subsection{Three-Dimensional Harmonic Confinement Potential}

In truth, the formulas derived before do not require any container
for the pressure field to be well-defined. In fact, we can use any
generic confining potential in its place, e.g. a three-dimensional
harmonic potential $V\!\left(\mathbf{r}\right)\!=\!\nicefrac{1}{2}m\omega^{2}r^{2}$.
Note that, despite allowing the gas particles to be located anywhere
in space, this potential naturally defines an effective volume given
by $V_{\text{eff}}\!=\!\left(2\pi/m\omega^{2}\beta\right)^{{\scriptscriptstyle 3/2}}$.
The equilibrium particle density is then shown to have a gaussian
radial profile, $n_{\text{eq}}\!\left(r\right)\!=\!(N/V_{\text{eff}})\exp\left[-(\nicefrac{3}{2})r^{2}/\xi^{2}\right]$,
with a characteristic width $\xi\!=\!\sqrt{3/m\omega^{2}\beta}$.
Even though there are no walls, a pressure field can still be defined
using Eq.\,\ref{eq:IdealGasLocal}, yielding

\vspace{-0.5cm}
\begin{equation}
P\left(r\right)=P_{0}\exp\left(-\frac{3r^{2}}{2\xi^{2}}\right)
\end{equation}
where $P_{0}\!=\!N/\beta V_{\text{eff}}$ is the pressure at the origin
and also the maximum value it can attain. Similarly, the buoyancy
force field is obtained

\vspace{-0.5cm}
\begin{equation}
\mathbf{B}\left(\mathbf{r}\right)=\frac{3P_{0}}{\xi^{2}}r\exp\left(-\frac{3r^{2}}{2\xi^{2}}\right)\hat{\mathbf{r}},
\end{equation}
being, unsurprisingly, directed in the outward radial direction. %

\vspace{-0.4cm}

\subsection{Centrifugal Buoyancy}

As a last example, we consider the nontrivial case of buoyancy inside
a fluid confined to a rotating cylindrical container of radius $\xi$
and length $L$. This system is assumed to be rotating around its
revolution axis, with a fixed angular velocity $\Omega$. This motion
induces a centrifugal force that is described by a potential $V\!\left(\rho\right)\!=\!-\nicefrac{1}{2}m\Omega^{2}\rho^{2}$,
where $\rho$ is the radial cylindrical coordinate. This inverse harmonic
potential is obviously not a confining one, but this is not a problem
as the later is provided by the cylinder's hard wall. In this case,
one can then define an effective volume, $V_{\text{eff}}\!=\!(2\pi L)/(\beta m\Omega^{2})\left[\exp\left(\nicefrac{1}{2}\beta m\Omega^{2}\xi^{2}\right)\!-\!1\right]$,
such that the equilibrium particle density is $n_{\text{eq}}\!\left(\rho\right)\!=\!(N/V_{\text{eff}})\exp\left(\nicefrac{1}{2}\beta m\Omega^{2}\rho^{2}\right)$
inside the cylinder, and $0$ outside (see Fig. \ref{fig.profiles}\,b).
Correspondingly, the buoyancy force field can be easily determined
as follows

\vspace{-0.4cm}

\begin{equation}
\mathbf{B}\left(\boldsymbol{\rho}\right)=-\frac{N}{V_{\text{eff}}}\beta m\Omega^{2}\rho\exp\left(\beta\frac{1}{2}m\Omega^{2}\rho^{2}\right)\hat{\boldsymbol{\mathbf{\rho}}},
\end{equation}
which is now directed inwards. As in the previous situations, the
natural tendency of a buoyant force is to balance existing the centrifugal
force acting upon any object floating inside. %

\vspace{-0.4cm}

\section{\label{sec:Conclusion-and-Outlook}Summary and Concluding Remarks}

\vspace{-0.4cm}

\noindent Many real-life physical situations, ranging from atmospheric
physics to buoyancy forces, can only be described by means of space-dependent
pressure fields. From a microscopic point-of-view, it is not trivial
to approach these thermodynamic concepts without introducing relatively
complex formulations that obstruct an efficient teaching. In this
paper, we have attempted to bypass this problem and provide a renewed
simple view on the microscopic definition of a pressure field inside
an ideal fluid that is acted upon by an arbitrary external force field.

For this purpose, we relied on the standard canonical formulation
of statistical mechanics, and studied a gas model with independent
classical particles moving on an external scalar potential $V\!(\mathbf{r})$.
In order to unambiguously define the local pressure field, an additional
short-ranged spherical potential was considered, so as to provide
the needed point-like barometer capable of probing pressure inside
the fluid. Using this setup, we have derived a closed-form formula
for the pressure field within the fluid, in the presence of arbitrary
external potentials, assuming the limit of very small probes (compared
to the variation scale of the external force field) with internal
potentials that approach impenetrable spherical walls. With this result,
we arrived at a local version of the ideal gas law, relating the value
of the pressure field in a given point, to the equilibrium density
of particles in that same point.

Finally, we retraced the same process to calculate the net force acting
upon the local barometer and, with it, derived a generalized version
of Archimedes' principle capable of describing buoyancy in non-homogeneous
force fields. We applied this formulation to three paradigmatic cases:
\textit{i)} a fluid under a uniform gravitational field, \textit{ii)}
a gas confined by a three-dimensional harmonic potential, and \textit{iii)}
a rotating two-dimensional fluid under a centrifugal force. By its
intrinsic richness but remarkable simplicity, we believe the formulation
presented here to be a relevant resource for teaching undergraduate
students the connections between statistical and thermodynamic descriptions
of fluids, allowing to consider situations that go beyond traditional
textbook material, but rather approach more realistic scenarios. Finally,
to assess the reaction of university level students, this microscopic
formulation of pressure fields was used as a basis for a problem posed
in the 2021 edition of the international Physics competition \textit{Plancks}\,\citep{plancks2021}.
The participants' responses included very physically rich and intuitive
discussions.

\vspace{-0.4cm}
\begin{acknowledgments}
The authors acknowledge financial support by the Portuguese Foundation
for Science and Technology (FCT) within the Strategic Funding UIDB/04650/2020
and COMPETE 2020 program in FEDER component (European Union), through
Project No. POCI-01-0145-FEDER-028887. S.M.J. and J.P.S.P. were further
supported by FCT Ph.D. grants PD/BD/142798/2018 and PD/BD/142774/2018,
respectively. The authors further thank Prof. J.M.B Lopes dos Santos,
J. Gonçalo Roboredo and J. Matos for fruitful discussions.
\end{acknowledgments}

\appendix

\vspace{-0.3cm}

\section{\label{sec:Appendix:DiracDelta}Useful Identities With Dirac-$\delta$
Functions}

In this section, we delve into more detail about the trick used for
the derivation of the local pressure and the buoyant force field in
Sect\,\ref{sec:Non-uniform-pressure} and \ref{sec:Microscopic-Derivation-of}.
Specifically, we want to show that

\begin{equation}
\frac{\partial e^{-\beta U_{\text{p}}}}{\partial r}\!=\!-\frac{\partial e^{-\beta U_{\text{p}}}}{\partial R}
\end{equation}
under reasonable assumptions. Recalling Eq.\,(\ref{eq:ProbePotential}),
we begin by noting that

\vspace{-0.5cm}
\begin{equation}
\frac{\partial U_{\text{p}}\left(r,R\right)}{\partial r}e^{-\beta U_{\text{p}}\left(r,R\right)}\!=\!-\frac{1}{\beta}\frac{\partial}{\partial r}e^{-\beta U_{\text{p}}\left(r,R\right)}.
\end{equation}
This function is highly localized around the surface of the sphere
$r=R$. Let's analyze how this object acts inside the integrals we
typically find:

\vspace{-0.5cm}

\begin{equation}
\int_{0}^{\infty}\!\!\!\!\!drg\left(r\right)\frac{\partial}{\partial r}e^{-\beta U_{\text{p}}\left(r,R\right)}
\end{equation}
where $g\left(r\right)$ is assumed to decay exponentially to zero
as $r$ increases. Such is the case when $g\left(r\right)=e^{-\beta V\left(r\right)}$.
Using integration by parts,

\vspace{-0.5cm}

\begin{align}
\int_{0}^{\infty}\!\!\!\!\!dr\left[\frac{\partial}{\partial r}e^{-\beta U_{\text{p}}\left(r,R\right)}\right]g\!\left(r\right) & \!=\!\!\\
\int_{0}^{\infty}\!\!\!\!\!\!dr\frac{\partial}{\partial r}\left[g\left(r\right)e^{-\beta U_{\text{p}}\left(r,R\right)}\right] & \!\!-\!\!\int_{0}^{\infty}\!\!\!\!\!\!dre^{-\beta U_{\text{p}}\left(r,R\right)}\frac{\partial}{\partial r}g\left(r\right),\nonumber 
\end{align}
the first term evaluates to zero because $g\left(\infty\right)\!=\!0$
by assumption and $e^{-\beta U_{\text{p}}\left(r,R\right)}\!=\!0$
when $r\!<\!R$ in the hard wall limit $U_{0}\!\rightarrow\!\infty$.
In this same limit, the second term simplifies to

\vspace{-0.5cm}

\begin{equation}
\!\!\!\int_{0}^{\infty}\!\!\!\!\!\!\!dre^{-\beta U_{\text{p}}\left(r,R\right)}\!\frac{\partial}{\partial r}g\!\left(r\right)\!=\!\!\!\int_{R}^{\infty}\!\!\!\!\!\!\!dr\frac{\partial}{\partial r}g\left(r\right)\!=\!-g\!\left(R\right)\!.
\end{equation}
Thus far, we have shown that

\vspace{-0.5cm}

\begin{equation}
\int_{0}^{\infty}\!\!\!\!\!drg\left(r\right)\frac{\partial}{\partial r}e^{-\beta U\left(r,R\right)}\!\!=\!g\left(R\right)
\end{equation}
for an arbitrary well-behaved function $g\left(r\right)$. Then, we
can say that $\nicefrac{\partial}{\partial r}\exp\left[-\beta U_{\text{p}}\!\left(r,R\right)\right]\!=\!\delta\left(r\!-\!R\right)$
under these conditions. Using the fact that $r\!=\!R$ under the action
of $\delta\left(r-R\right)$, we get that

\vspace{-0.5cm}
\begin{equation}
\frac{\partial e^{-\beta U_{\text{p}}}}{\partial r}\!=\!\delta\left(r\!-\!R\right)\!=\!\delta\left(r\!-\!R\right)\frac{R}{r}\!=\!\frac{R}{r}\frac{\partial e^{-\beta U_{\text{p}}}}{\partial r}.
\end{equation}
Finally, using the fact that $R(\partial U_{\text{p}}/\partial r')\!=\!-r^{\prime}(\partial U_{\text{p}}/\partial R)$,
the original assertion,

\vspace{-0.5cm}

\begin{equation}
\frac{\partial e^{-\beta U_{\text{p}}}}{\partial r}\!=\!-\frac{\partial e^{-\beta U_{\text{p}}}}{\partial R}
\end{equation}
immediately follows.

\vspace{-0.5cm}

\section{\label{sec:Appendix:-Density-of}Equilibrium Density of particles
in the Presence of an External Field}

In this appendix, we derive the equilibrium density of particles for
an ideal classical gas under the influence of an arbitrary scalar
potential $V\!(\mathbf{r})$. When deriving the expressions of the
pressure and buoyancy force fields, this is a necessary quantity that
appear in all the expressions. To be precise, this quantity is found
by evaluating the statistical average of the following characteristic
function

\vspace{-0.5cm}

\begin{equation}
n\!\left(\mathbf{r}\right)\!=\!\!\sum_{i=1}^{N}\delta\left(\mathbf{r}\!-\!\mathbf{r}_{i}\right)
\end{equation}
Considering only the gas subject to an external potential (with no
barometer present), its average in the canonical ensemble is

\vspace{-0.4cm}

{\small{}
\begin{equation}
n_{\text{eq}}\!\left(\mathbf{r}\right)\!=\!\frac{\int\!\!d^{{\scriptscriptstyle \overset{\overset{\text{(3N)}}{}}{\,}}}\!\!\!\mathbf{r}^{\prime}\!\!\int\!\!d^{{\scriptscriptstyle \overset{\overset{\text{(3N)}}{}}{\,}}}\!\!\!\mathbf{p}\sum_{i}\!\delta\left(\mathbf{r}\!-\!\mathbf{r}_{i}\right)e^{-\beta\sum_{j}\left(\frac{\mathbf{p}_{j}^{2}}{2m}+V\!(\mathbf{r}_{j})\right)}}{\int\!\!d^{{\scriptscriptstyle \overset{\overset{\text{(3N)}}{}}{\,}}}\!\!\!\mathbf{r}^{\prime}\!\!\int\!\!d^{{\scriptscriptstyle \overset{\overset{\text{(3N)}}{}}{\,}}}\!\!\!\mathbf{p}e^{-\beta\sum_{j}(\frac{\mathbf{p}_{j}^{2}}{2m}+V(\mathbf{r}_{j}))}}
\end{equation}
}which reduces to

\vspace{-0.4cm}
\begin{equation}
n_{\text{eq}}\!\left(\mathbf{r}\right)\!=\!\frac{N}{V_{\text{eff}}}\exp\left(-\beta V\left(\mathbf{r}\right)\right)
\end{equation}
and so the particles are more likely to be found in regions where
the external potential $V$ is smaller. In equilibrium, this is expected
intuitively. If $V\!\left(\mathbf{r}\right)$ is constant, then the
density of particles reduces to the familiar expression $N/V$.

\[
\]

\[
\]

\bibliographystyle{plain}
\bibliography{References}

\end{document}